\documentclass[aps,twocolumn,showpacs]{revtex4}
\usepackage{graphicx,color}

\begin{document}
\draft \title{Comment on "Motion of an impurity particle in an ultracold quasi-one-dimensional gas of hard-core bosons [Phys. Rev. A 79, 033610 (2009)]"}
\author{S. Giraud and R. Combescot} \address{Laboratoire de Physique Statistique, Ecole Normale Sup\'erieure, UPMC Paris 06, 
Universit\'e Paris Diderot, CNRS, 24 rue Lhomond, 75005 Paris,  France}
\begin{abstract}
Very recently Girardeau and Minguzzi [arXiv:0807.3366v2, Phys. Rev. A 79, 033610 (2009)] have studied an impurity in a one-dimensional gas of hard-core bosons. In particular they deal with the general case
where the mass of the impurity is different from the mass of the bosons and the impurity-boson interaction is not necessarily infinitely repulsive.
We show that one of their initial step is erroneous, contradicting both physical intuition and known exact results. Their results in the general case apply
only actually when the mass of the impurity is infinite.
\end{abstract}

\maketitle

The subject of mixtures of ultracold gases is of very high current interest \cite{gps}, as more and more experimental results
are coming out of laboratories aiming at investigating this almost unexplored quantum domain. An extreme case
of mixtures is a single foreign atom (an "impurity") immersed in a sea of identical atoms. This system is quite
attractive because, despite its apparent simplicity, it is still fairly complex and moreover it is an excellent starting
point to deal with more complicated situations \cite{lrgs}. As usual one-dimensional cases lend themselves to much easier
numerical handling and possibly to analytical solutions which makes them worth studying in details.

Quite recently Girardeau and Minguzzi \cite{gm} (GM) have studied an impurity in the presence of a 1D gas of hard-core bosons.
Similarly these bosons interact with the impurity by a contact potential with finite strength. As it is well known, when the boson-boson
interaction is extremely strong (or the boson gas very dilute) making the bosons impenetrable, the system enters the so-called Tonks-Girardeau regime \cite{tg}
where, except for changes of sign, the wave function of the bosons is identical to the one of free fermions, the Pauli principle taking care
that two bosons are not in the same location. When the mass of the impurity is equal to the mass of the bosons, Girardeau and Minguzzi
have shown that the same kind of arguments allow to write the wave function of the total system.

However in a next step they aim to generalize the system under study to the case where the impurity does not have the same
mass as the bosons and the boson-impurity interaction is not infinitely strong. After a canonical transformation which makes completely explicit the transformation 
from a boson problem to a fermion problem, they have the following starting Hamiltonian ${\hat H}_F$ (Eq.(3) of \cite{gm}) for free fermions of mass $m$ (this fermion mass is actually taken as unity by GM) interacting with an impurity of mass $m_i$:
\begin{eqnarray}\label{}
{\hat H}_F \!=\! \! \!\int \!\!dx \;{\hat \psi}^{\dag}_F(x)\!\left[-\frac{1}{2m}\frac{\partial^2}{\partial x^2}\right]\!{\hat \psi}_F(x)
\!-\! \frac{1}{2m_i} \frac{\partial^2}{\partial y^2}\! +\! \lambda \,{\hat \rho}_F(y)
\end{eqnarray}
Here ${\hat \psi}_F(x)$ is the annihilation field operator for a fermion at $x$, ${\hat \rho}_F(y)={\hat \psi}^{\dag}_F(y){\hat \psi}_F(y)$ is the fermion density operator at the impurity position $y$ and $\lambda$ the strength of the fermion-impurity interaction. This Hamiltonian is written in a mixed representation, first quantization for the impurity and second quantization for the fermions. If we write it fully in first quantization, we have:
\begin{eqnarray}\label{hmcg}
{\hat H}_F =   -\frac{1}{2m}\,\sum_{j=1}^{N}\frac{\partial^2}{\partial x_j^2}
- \frac{1}{2m_i} \frac{\partial^2}{\partial y^2} + \lambda \,\sum_{j=1}^{N} \delta(x_j-y)
\end{eqnarray}
This expression is equivalent to the one written explicitly by McGuire \cite{mcg}. Instead, if we write ${\hat H}_F$ fully in second quantization, 
and make use of annihilation $c_{k}$ and creation $c^{\dag}_{k}$ operators for the fermions in plane wave states $k$, related to the above field operator by ${\hat \psi}_F(x)= \sum_{k}e^{ikx}\,c_{k}$, we obtain:
\begin{eqnarray}\label{gch}
{\hat H}_F=\!\sum_{k}\epsilon _{k}c^{\dag}_{k}c_{k}+\!\sum_{q}E(q)b^{\dag}_{q}b_{q}
+\lambda\!\!\sum_{k k' q q'}\!\!\delta_{k k' q q'}c^{\dag}_{k}c_{k'}b^{\dag}_{q}b_{q'}
\end{eqnarray}
where $b_{q}$ and $b^{\dag}_{q}$ are the corresponding annihilation and creation operators for the impurity, while $\epsilon _k=k^2/2m$ and $E(q)=q^2/2m_i$ are respectively the fermion and the impurity kinetic energy. The Kronecker symbol $\delta_{k k' q q'}$ ensures momentum conservation $k+q=k'+q'$ in the fermion-impurity scattering. This last form Eq.(\ref{gch}) is the one we have written in our very recent investigation \cite{gc} of this same problem.

In their paper GM perform a Lee, Low and Pines canonical transformation, which in the present case amounts to a translation on the fermions coordinates to shift the position of the impurity at the origin. In this way they end up with the following Hamiltonian:
\begin{eqnarray}\label{gm1}
{\hat H}_F =  \int dx \;{\hat \psi}^{\dag}_F(x)\left[-\frac{1}{2m}\frac{\partial^2}{\partial x^2}\right]{\hat \psi}_F(x) \\  \nonumber
+ \lambda \,{\hat \rho}_F(0)+\frac{{\hat p}_F^2}{2m_i}-\frac{q{\hat p}_F}{m_i}+\frac{q^2}{2m_i}
\end{eqnarray}
where $q$ is the total momentum of the system, whose introduction allows to get rid of the impurity momentum.

Then GM argue that the term ${\hat p}_F^2/2m_i$ in this Hamiltonian is negligible in the thermodynamic limit. Our view is that this last step does not make any sense physically, and is in plain contradiction with known exact results on this problem.

The physical problem is the most striking in the case where the total momentum of the system is zero $q=0$ (the ground state of the system is clearly found in this case). If we accept GM statement, only the first two terms in the right-hand side of Eq.(\ref{gm1}) are left. This implies that the detailed physics of the system is completely independent of the mass of the impurity $m_i$. In particular a very heavy impurity (which behaves as a mere scattering center) and a very light impurity would give exactly the same quantitative physical results. Consideration of the special case of infinitely strong repulsion $\lambda \rightarrow \infty$ makes this conclusion completely unacceptable. Indeed, going back to the simple form Eq.(\ref{hmcg}) for the Hamiltonian, a very heavy impurity would merely act as a fixed infinitely repulsive boundary for the fermions located at $y$. By contrast the very light impurity will very strongly push away the fermions in order to be widely delocalized and to reduce in this way its large kinetic energy arising from Heisenberg uncertainty principle. Clearly the energy required to put the impurity in the Fermi sea (that is the impurity chemical potential) is quite different in these two cases since the physical modification of the Fermi sea brought by the impurity is quite different. Naturally these qualitative physical considerations are fully born out by the quantitative exact results to which we turn now.

The case where the impurity mass is equal to the fermion mass $m_i=m$ has been completely solved analytically by McGuire \cite{mcg} for any value of the coupling strength $\lambda$, with a technique which is a Bethe ansatz method. In this way he has calculated the impurity chemical potential, as well as its effective mass. Let us concentrate on the impurity chemical potential $\mu _i$, which is just the opposite of the impurity binding energy: $E_b=-\mu _i$. This is the easiest quantity to understand physically. Unfortunately GM have only calculated the fermion distribution around the impurity $\rho(x-y)$, not the binding energy. The result of McGuire is :
\begin{eqnarray}\label{muex}
\frac{E_b}{E_F}&=&-\frac{2}{\pi}\left[y-\frac{\pi}{2}y^2+(1+y^2)\arctan y\right]
\end{eqnarray}
where $y=m\lambda/2k_F$, $E_F=k_F^2/2m$ is the Fermi energy of the fermions and $k_F=\pi n$ the Fermi wave vector of these fermions with density $n$.

On the other hand we have also solved analytically \cite{gc} the much simpler problem where the impurity mass is infinite $m_i=\infty$. In this case, as we have indicated above, the impurity is just a fixed scattering center for the fermions and one has just to deal with the one-body problem of a fermion in the presence of the scattering potential. Our result is:
\begin{eqnarray}\label{muinf}
\frac{E_b}{E_F}=-\frac{1}{\pi}\left[z-\frac{\pi}{2}z^2+(1+z^2)\arctan z\right]
\end{eqnarray}
where $z=m\lambda/k_F$. Although quite similar to Eq.(\ref{muex}) (there is just a factor 2 in the definition of $z$, as compared to $y$, and also a factor 2 in the $E_b/E_F$ expressions if we forget the difference between $y$ and $z$), this result is different. This shows explicitly that the physical results obtained from ${\hat H}_F$ depend on the mass of the impurity, even when the total momentum of the system $q$ is zero, since they are different for $m_i=m$ and for $m_i=\infty$.

Turning now to the physical quantity calculated by GM, namely the impurity-fermion distribution function (i.e., from the boson to fermion mapping, the impurity-boson distribution function in their original problem), our interpretation is that their calculations correspond only to the $m_i=\infty$ case (since it is in this case true that the term ${\hat p}_F^2/2m_i$ disappears in the Hamiltonian). We have also calculated \cite{gc} this quantity in this case and found an analytical expression under the form of a simple integral. It has been plotted in our Fig.~13 for various values of the coupling constant (not all the same as GM). We have checked that our results are the same as those of GM if we take the same values of the coupling constant. In particular when $\lambda=\infty$ we obtain for the normalized distribution $\rho(x)=1-\sin(2k_Fx)/(2k_Fx)$, in agreement with their result since $j_0(z)=\sin (z)/z$. On the other hand McGuire \cite{mcg} has also calculated this distribution function in the case of equal masses $m_i=m$ and shown plots of it for several values of the coupling constant. A particularly striking feature (see the remark Ref.(13) in \cite{gc}) of his results is that $\rho(x) \leq 1$ in the repulsive case, while $\rho(x) \geq 1$ in the attractive one. This is in striking contradiction with GM results. In particular McGuire obtains in the infinitely repulsive case $\lambda=\infty$ that $\rho(x)=1-\left[\sin(k_Fx)/(k_Fx)\right]^2$, which is different from the above result for $m_i=\infty$. Actually, as noted by McGuire, this last result can be deduced immediately from Girardeau's work \cite{tg}, since in this case the impurity behaves just as an additional fermion. More generally if one were to apply the valid treatment of GM section II, which deals with the case $m_i=m$ with infinitely strong impurity-fermion repulsion, one would discover that the results are in contradiction with those of section IV.

Finally let us comment briefly on the arguments presented by GM in their section V to justify their omission of the ${\hat p}_F^2/2m_i$ term. They state that the average of this term is ${\mathcal O}(1)$. This is obviously correct (since, as they notice at the end of this section, it is directly related to the fluctuation of the impurity momentum, which is clearly ${\mathcal O}(1)$). Hence it is definitely negligible in the thermodynamic limit compared to the total ground state energy, which is merely the energy of a free Fermi sea $E_0=L k_F^3/(6\pi m)=NE_F/3$ with $n=N/L=k_F/\pi $. However the physics of ${\mathcal O}(L)$ is completely trivial and uninteresting because it is the physics of a free Fermi sea.
All the interesting quantities in this problem, related to the physics of the impurity, are of order ${\mathcal O}(1)$, just as the binding energy given by Eq.(\ref{muex}) or Eq.(\ref{muinf}). Accordingly one has to keep all the ${\mathcal O}(1)$ quantities in order to deal properly with the impurity problem. The final (unsubstantiated) suggestion that the contribution to $\rho(x)$ could be ${\mathcal O}(L^{-1})$ is disproved by the explicit examples given in the preceding paragraph.

We acknowledge two mail exchanges on this matter with M. D. Girardeau, at the stage where the work was
posted on ArXiv, in which we tried to convince him that there was a problem. Unfortunately we were unsuccessful.

\end{document}